# Dynamic Data Consistency Tests Using a CRUD Matrix as an Underlying Model

Miroslav Bures
FEE, CTU in Prague
Karlovo nam. 13, 121 35 Praha 2
Czech Republic
miroslav.bures@fel.cvut.cz

Vaclav Rechtberger
FEE, CTU in Prague
Karlovo nam. 13, 121 35 Praha 2
Czech Republic
rechtva1@fel.cvut.cz

## ABSTRACT

In testing of software and Internet of Things (IoT) systems, one of necessary type of tests has to verify the consistency of data that are processed and stored in the system. The Data Cycle Test technique can effectively do such tests. The goal of this technique is to verify that the system processes data entities in a system under test in a correct way and that they remain in a consistent state after operations such as create, read, update and delete. Create, read, update and delete (CRUD) matrices are used for this purpose. In this paper, we propose an extension of the Data Cycle Test design technique, which is described in the TMap methodology and related literature. This extension includes a more exact definition of the test coverage, a reflection of the relationships between the tested data entities, an exact algorithm to select and combine read and update operations in test cases for a particular data entity, and verification of the consistency of the produced test cases. As verified by our experiments, in comparison to the original Data Cycle Test technique, this proposed extension helps test designers to produce more consistent test cases that reduce the number of undetected potential data consistency defects.

## 1. INTRODUCTION

Under the current practices of software testing, an established set of test case design techniques is used, which exercise different aspects of the system under test (SUT). The most common approaches focus on the exercising of paths in SUT processes, transitions of SUT states or consistency of data objects processed by SUT [1-3]. In this paper, we focus on testing the consistency of data objects. This aspect is important for integration and end-to-end (E2E) testing of various systems working with data of nontrivial complexity. To give some examples, complex enterprise software systems of various IoT systems are typical fields where consistency of processed data has to be treated properly, considering current reliability challenges in the field of software and IoT [4].

In this study, we extend the Data Cycle Test (DCyT), which is also known as the CRUD test [3,5,6,7]. The DCyT is based on a CRUD matrix. As a reference point, we use the definition of the method given in TMap Next [3]. We propose an extension to this technique, which provides better control to the test coverage and ensures better efficiency of the test cases that are produced. Which gives the tester better control over the test coverage and improves efficiency of the test cases produced by the proposed technique.

The data objects processed by the SUT are defined by the data entities: a **data object** is an instance of a particular data entity. A **data entity** $e$ consists of a set of attributes and possibly of other data entities:

$e = (A, D)$, $A$ is a set of attributes, $A$ contains at least one attribute. $D$ is a set of data entities. $D$ can be empty. For the purpose of our proposal, the data entities are identified on a conceptual level of the design. The specific physical representation in the database is a detail that, generally, we do not focus on.

Each of the data objects is created, read, updated or deleted by a particular SUT function. The flow of these operations creates the lifecycle of the data object. Obviously, this type of data object lifecycle would not be performed completely inside of one particular process; instead, its functions may be performed within a number of processes.

A number of defects can be caused by incorrect handling of data objects by the SUT. In the discussed technique, we are interested in the following situation:

Function $f_1$ represents an abstracted unit of SUT functionality affecting the data objects. The function $f_1$ changes the values of data object attributes. A data object $o$ with defined attributes $A = \{a_1, \ldots, a_n\}$ has values of these attributes $V = \{v_1, \ldots, v_n\}$. After being processed by the function $f_1$, the data object, according to the SUT specification, should have values of these attributes $V' = \{v_1', \ldots, v_n'\}$. If a defect is present in function $f_1$, the attribute values of data object $o$ can be set to $V'' = \{v_1'', \ldots, v_n''\}$. When $V' \neq V''$, we define the **data object as inconsistent**. When $o$ is handled by other SUT functions, this can cause faulty behavior of some of these functions.

A data object can also be made inconsistent by a Create operation. This can occur in the following scenario. Function $f_1$ creates a data object $o$. By the specification of the SUT, this created data object $o$ should have the values of its attributes $V' = \{ v_1', …, v_n' \}$. A defect in function $f_l$ causes the actual values of these attributes to be $V'' = \{ v_1'', …, v_n'' \}$. If $V' \neq V''$, the data object $o$ was created as inconsistent. Also, in this situation, faulty behavior can arise in other SUT functions when $o$ is handled by them.

The **SUT function** (or function) is a feature of an SUT that performs any Create, Update, Read and Delete (C, R, U, D) operation on a data object. In this proposal, we do not take into account any features of the SUT which do not perform any of the C, R, U, D operations. The level of detail and granularity of the functions depends on the particular case and the experience of the test designer. Nevertheless, the discussed technique is not dependent on the granularity of the data entities and their functions.

When discussing the inconsistency of a data object, we do not mean a data consistency in a relational database sense. Even when all database schema constraints are fulfilled, the particular values of a data object can still be inconsistent from a semantic point of view, as we have explained above.

The paper is organized as follows. Section 2 provides a more detailed description of DCyT. Here, we also discuss its limitations and areas in which we propose extension of this technique. Then, the EDCyT (our extension of DCyT) is presented in Section 3. Section 4 provides details about the conducted experiments. An Overview of related work is provided in Section 5. The last section concludes the paper.

## 2. THE DATA CYCLE TEST

In this chapter, we introduce the common definition of Data Cycle Test (DCyT), which is presented for example by [3,5-7]. Then, we discuss its limitations and introduce areas, in which we extend this method to Extended Data Cycle Test (EDCyT).

### 2.1 Basic Concepts

First, let us introduce some terminology, which is used in the following text.

The data entity, data object, inconsistent data object and SUT function has been defined in Section 1 already.

$F = \{f_1, …, f_n\}$ is a set of all the SUT functions, and $E = \{e_1, …, e_p\}$ is a set of all the data entities that were taken into account for the test design.

Then, the **CRUD matrix** is defined as $\mathbf{M} = (m_{i,j})_{n,p}$, $n = |F|$, $p = |E|$, $m_{i,j} = \{ o \mid o \in \{ C, R, U, D \} \} \Leftrightarrow$ function $f_i \in F$ performs the respective Create, Read, Update or Delete operations on the data entity $e_j \in E$ }.

### 2.2 The DCyT Principle

The common presentation of the Data Cycle Test (DCyT) technique [3,5,6,7] uses a CRUD matrix **M** as a model of the SUT.

Output of the DCyT is a set of test cases created to verify consistency of the data objects during their lifecycle in the SUT, in particular during the C, R, U, D operations, which may be performed on these entities. For each data entity, one test case is created typically.

A **test case** $c$ for data entity $e$ is a sequence of **test steps** $\{s_1, .., s_n\}$. Each of these steps is a pair $(f_s, o_s)$, and $f_s \in F$, $o_s \in \{C, R, U, D\}$ is an operation that is performed on the data entity $e$ by the function $f_s$. Possible functions $f_s \in F$ in the test case $c$ are selected in accord to the CRUD matrix **M**.

Test case $c$ is started by the Create operation. Then, all possible Update operations follow. Finally, $c$ ends by a Delete operation. In the test case, a Read operation is performed after every Create, Update and Delete operation.

TMap DCyT does not define the test coverage criteria exactly. However, two levels of the test coverage are discussed:

(1) After each Create, Update or Delete operation, one Read operation is performed (a lightweight test coverage). Which Read operation has to be selected is not defined in the method, and

(2) After each Create, Update or Delete operation, all Read operations for entity $e$ are carried out once (an intense test coverage; however, leading to high number of test steps, which even may seem as not realistic to be executed in a real industry software testing project).

As a simple completeness check, TMap states, that all C,R,U,D operations with data entities of the CRUD matrix performed by the SUT functions shall be present in the final set of the test cases.

### 2.3 Opportunities to Extend the DCyT

We identify four areas, in which efficiency of DCyT can be increased:

(1) Exact test coverage criteria have to be defined to be able to regulate intensity of test cases produced by the technique (and thus time required to perform the tests). This is key issue for having test coverage under control and for planning of test effort. In DCyT, coverage criteria are described in a vague form only.

(2) Exact algorithm, how to select and combine read and update operations in test cases for the particular data entity is not presented.

(3) Also, an exact method, how to ensure that produced test cases are consistent (test case can be executed in the SUT) is not presented.

(4) Data dependencies and influencing behavior between tested data entities are not reflected in the test case creation process. Only case of master and derived data is discussed briefly [3], but this concept does not cover all cases of data dependencies, which can occur in SUT.

In this paper, we are covering these issues by proposal of the Extened Data Cycle Test (EDCyT).

## 3. EXTENSION OF THE DCyT

The proposed extension of the DCyT technique consists of a set of enhancements in two areas: (1) extension to the CRUD matrix and (2) preparation of the test cases for dynamic testing.

When referencing a "standard CRUD matrix", we use a CRUD matrix with the operations C, R, U, D, defined as **M** above. In addition, we use the abbreviation EDCyT to describe a proposed extension of DCyT.

### 3.1 Extension of the CRUD Matrix

In the following section, we propose several enhancements to a CRUD matrix, which will be used further on in design of the test cases

### 3.1.1 Reflection of data entity attributes in the R and U operations

In a test case created by the DCyT, an U operation shall be followed by one or more R operations. Is this criterion specific enough to produce efficient test cases?

Consider the following case. The test case $c$ for data entity $e$ is a sequence of test steps $\{s_0, .., s_x, s_y, ..., s_n\}$. Test step $s_x = (f_x, \{U\})$. Test step $s_y = (f_y, \{R\})$. Data entity $e$ consists of a set of attributes $A$. The function $f_x$ performs an U operation and updates a set of attributes $A_x \subset A$. The function $f_y$ then perform a R operation and updates a set of attributes $A_y \subset A$. When $A_x \cap A_y = \emptyset$, by the R operation we do not test any of the attributes changed by the U operation. Instead of $s_y$, we need to use a $s_z = (f_z, \{R\})$, where function $f_z$, perform a R operation on a set of attributes $A_z \subset A$, where ideally $A_x = A_z$, or at least $A_x \cap A_z \neq \emptyset$.

To ensure this, we need to reflect the attributes when combining flow of U and R operations in the test cases. First of all, information which attributes are handled by particular operations has to be captured in the CRUD matrix.

Let's extend the definition of the CRUD matrix already given in Section 2.1.

The **CRUD matrix** is $M=(m_{i,j})_{n,p}$, $n = |F|$, $p = |E|$, $m_{i,j} = \{ o \mid o \in \{$ 'C', $r$, $u$, 'D' $\} \Leftrightarrow$ function $f_i \in F$ performs the respective Create, Read, Update or Delete operations on the data entity $e_j \in E$ $\}$. $r = ($ 'R', $A_R)$. $A_R \subseteq A_j$ is set of attributes read by the R operation. $u = ($ 'U', $A_U)$. $A_U \subseteq A_j$ is set of attributes update by the U operation. $A_j$ is set of attributes of the data entity $e$. 'C', 'R', 'U', 'D' are symbols defining that function $f_i \in F$ performs Create, Read, Update or Delete operations on the data entity $e_j \in E$.

### 3.1.2 Operation influenced

We extend the set of C, R, U, D operations that are used in a standard CRUD matrix by a special type of operation called **Influenced**, which is denoted as I($e$), $e \in E$. The test designer can use this operation to capture the following situations, which are relevant for the creation of the test cases:

Data entity $e_1 \in E$ is **influenced** (denoted as I($e_2$)) by function $f \in F$ if the function $f$ uses a data entity $e_2 \in E$ through one or more of the C, U, D operations, and the particular data content of entity $e_1$ is changed as a consequence of function $f$, which primarily uses the entity $e_2$. In a CRUD matrix, the situation described above is captured by operation I($e_2$) in the cell for function $f$ and entity $e_1$.

The test designer can use the operation I to capture important cases of influencing entities from the test design point of view. By adding operation I, we are able to distinguish, which influencing data entities would be reflected in the test cases and which would not (the algorithm for definition of the test cases follows in chapter 3.4.1).

For test design purposes, it is not necessary to capture all the facts that must be reflected by the operation Influenced in the matrix. As we will describe later, this enhancement is also functional in situations where only part of the operations Influenced are captured in the CRUD matrix.

Cannot this situation, for which we are using I($e$), be described by only capturing the C, R, U, D operations?

TMap Next considers a similar case; in particular, "Derived data with processing functions" [3]. Data entities are categorized according to master data and derived data. If the derived data are produced as a result of a specific business process, in which master data are used, the situation should be captured by adding an R operation into the master data.

Although these situations look similar at first glance, this is not the same situation that I(e) is capturing. In contrast to TMap situation, by using I($e$), we are able to capture the following case: "The SUT function $f$ is modifying a particular data entity $e_2$, and this is also influencing (exactly) the $e_1$ data entity". We need this type of information to design more efficient and accurate test cases, especially for the situations, when the test designer needs to prioritize and select influencing entities, which are important in the test design. Without the operation I, adding R to master data as suggested in TMap Next description of the technique [3] could lead to unnecessary test case steps for low priority cases.

The I($e_2$) operation can be identified by following ways during the test design:

1. Information in test basis (business or technical specification), that particular function $f$ using a data entity $e_2 \in E$ through one or more of the C, U, D operations also changes another data entity $e_1$
2. Conceptual database schema if prepared by technical designer of SUT
3. Physical database schema of SUT
4. Analysis of SUT source code operating with data entities when Object-Relational Mapping (ORM) framework is used
5. Experience of testers with knowledge of the SUT

All of the possible methods can be performed manually. Methods 3 and 4 can be potentially automated. Using Physical database schema of SUT, master-detail relationships can be analyzed automatically and used for definition of Influenced operations for relevant data entities. Also getting this information by analyzing of the source code in case of Object-relational mapping (ORM) framework can be potentially automated.

### 3.1.3 Operation best read

The operation Best read helps the test designer to select suitable Read operations when constructing a test case. This operation is denoted by the letter B in the CRUD matrix. B is a Read (R) operation that is performed by function $f \in F$ on entity $e \in E$, and it fulfills minimally one of the following conditions:

1. $A$ is the set of attributes of the data entity $e$. Function $f$ reads and displays the biggest subset $A' \subseteq A$ of the data entity attributes.
2. Function $f$ is the most frequent read function that is performed on data entity $e$;
3. Function $f$ is the most important read function that is performed on the data entity $e$ from a test prioritization point of view;
4. The code for reading data entity $e$, which is used in the function $f$, is also employed by other sets of functions $F' \subset F$ that perform a read operation on data entity $e$.

Practically, a B operation replaces an R operation in a CRUD matrix, which fulfills the criteria described above. For each entity $e \in E$, the test designer selects one of the read operations that are performed on the entity *e* as the best read, based on current information.

Let's compare the operation Best read to the simple exercise of one randomly selected read operation, which is a default approach presented in TMap Next [3]: by using the operation Best read, we may set the test coverage of the read operations to specific situations where the detection of more significant defects are likely to occur. We will address this issue later on when we explain how this operation will be used in the design of test cases.

In case we can't identify any of R operations as suitable candidate for B operation for particular data entity *e*, B operation could be performed as direct SQL query to the SUT database to verify the state of the data entity.

## 3.2 Generation of the Dynamic Test Cases

As we have already pointed out, in common definition of DCyT, the test coverage aspect is not formally defined nor discussed. Also a guide of how to select and to combine the operations is not presented at a level that is sufficient for the test design process. In the proposed extension, our goal is to consider these points.

Let us start with a recommendation for the test coverage criteria that are necessary for an Extended Data Cycle Test, which will also determine the method, and how the dynamic test cases are created.

### 3.2.1 Extension of test cases to reflect data entity attributes

In the EDCyT, we extend the test case steps by the particular attributes, which are modified and read by Update and Read operations.

The **test case** *c* for data entity *e*∈*E* is a sequence of **test steps** {$s_1$, .., $s_n$}, where each of these steps is a pair ($f_s$, $o_s$), and $f_s \in F$, $o_s \in$ {'C', *r*, *u*, 'D'}. *r* = ( 'R', $A_R$ ). $A_R \subseteq A$ is set of attributes read by R operation in the test step. *u* = ( 'U', $A_U$ ). $A_U \subseteq A$ is set of attributes updated by U operation in the test step. *A* is set of attributes of the data entity *e*. 'C', 'R', 'U', 'D' are symbols defining that function $f_s \in F$ performs Create, Read, Update or Delete operations on the data entity *e*. Possible functions $f_s \in F$ in the test case *c* are selected by the CRUD matrix **M**.

### 3.2.2 Test coverage criteria

As previously introduced, a test case is created for a particular entity *e*∈*E* as a sequence of the C, R, U, D, and B operations that are performed by the corresponding functions $f_1..f_n \in F$ on an entity *e*. For each entity *e*∈*E*, a test case using different test coverage criteria can be created.

Adding the extension of the CRUD matrix that was presented above, six levels of test coverage can be introduced:

**0R - Simple read**

Sequence of C, U, D operations on an entity *e*, each of C, U, D operations is followed by R operation on entity *e* selected subjectively by the test designer in accord to the knowledge about priority. If this knowledge is not available, R operation is selected randomly.

**0B - Simple best read**

Sequence of C, U, D operations on an entity *e*, each of C, U, D operations is followed by B operation on entity *e*.

**1R - All reads one time**

Sequence of C, U, D operations on an entity *e*, each of C, U, D operations is followed by a one or more R operations on entity *e*.

Each of the R operations on entity *e* is used once. This includes the B operation (as B is a special type of R operation).

**IF** ((total count of C, U, D operations on entity e) < (total count of R operations on entity e)) **THEN** distribute R operations between particular C, U and D operations to keep similar count of R after each C, U and D.

**IF** ((total count of C, U, D operations on entity e) > (total count of R operations on entity e)) **THEN** follow the remaining U, D operations by B operation.

**1RI - All reads one time, I operations reflected**

Test case created by the coverage type 1R, extended by operations included in the test case by the following algorithm:

*e* is the data entity for which the test case is being created

$F'$ = { $f'$ | $f' \in F$

$\wedge$ (line in the CRUD matrix corresponding to the function $f'$ contains operation I(*e*))

$\wedge$ ($f'$ is executed as a part of test case created for entity *e* by the coverage type 1R) }

**FOR EACH** $f' \in F'$ {

   $E'$ = { $e'$ | $e' \in E$

   $\wedge$ (column in the CRUD matrix corresponding to the entity *e'* contains operation I(*e*), performed by function $f'$) }

   **FOR EACH** $e' \in E'$ {

      **FOR EACH** of C, U, D operations on entity *e* performed by function $f'$ {

         Follow the C, U, D operation on entity *e* performed by function $f'$ by B operation on entity *e'*. If B operation is not defined for entity *e'*, use an R operation of entity *e'* instead: select this operation subjectively by the test designer in accord to the knowledge about priority. If this knowledge is not available, select an R operation randomly.

      }

   }

}

**NR - All reads after all changes**

Sequence of C, U, D operations on an entity *e*, each of C, U, D operations is followed by all R operations on entity *e*.

**NRI - All reads after all changes, I operations reflected**

Test case created by the coverage type NR, extended by operations included in the test case by the same algorithm as defined in the 1RI coverage criteria above.

Generally, in the variants 1RI and NRI, we are not only testing the data entity e, for which the test case was created, but we also performed R operations on the other data entities that were influenced by the functions exercised on data entity *e*. The goal was to verify that the other influenced data entities were in a consistent state, and they were changed according to the specifications of the functions exercised on entity *e*.

When using a standard CRUD matrix, two levels of test coverage, 0R and NR, are applicable.

### 3.2.3 Consistency and efficiency of the test cases

In the TMap description of the DCyT method, two important point is not specifically addressed: namely,

(1) which sequences of functions $f \in F$ are actually allowed to be performed in a SUT, and
(2) creating a test case for a data entity $e$, which combinations of U and R operations make sense to be executed in a sequence (regarding the fact that these U and R operations can change and read different subsets of attributes of the data entity $e$)?

Such situations can invalidate the test cases created by the DCyT. This applies also on the EDCyT, regardless of the test coverage level.

The **test case** $c = \{ s_1, \ldots, s_n \}$ for data entity $e$ is **inconsistent**, if it contains two test steps $s_a$ and $s_b$, $s_a$ is preceding $s_b$, when two subsequent C, R, U, D, and B operations performed by functions $f_a \in s_a$ and $f_b \in s_b$ in the test case cannot be performed in the SUT because we cannot reach a proper SUT state to execute the function $f_2$ from the SUT state reached by function $f_1$.

To repair an inconsistent test case, we insert a sequence $s_{c1}, \ldots, s_{cn} \in F$ between the steps $s_a$ and $s_b$. Adding the functions $f_{c1}, \ldots, f_{cn}$, $f_{c1} \in s_{c1}$, $f_{cn} \in s_{cn}$ allows to reach a proper SUT state to execute the function $f_2$ from the SUT state reached by function $f_1$.

Next, we have to prevent an inefficiency of the test cases, caused by the situation, when U operation is followed by R operation which is not actually testing changed attributes.

The following case can occur. The test case $c$ for data entity $e$ is a sequence of test steps $\{s_0, \ldots, s_x, s_y, \ldots, s_n\}$. Test step $s_x = (f_x, (\text{'U'}, A_x))$. Test step $s_y = (f_y, (\text{'R'}, A_y))$. Data entity $e$ consists of a set of attributes $A$. The function $f_x$ performs an U operation and updates a set of attributes $A_x \subset A$. The function $f_y$ then perform a R operation and updates a set of attributes $A_y \subset A$. When $A_x \cap A_y = \emptyset$, by the R operation we do not test any of the attributes changed by the U operation. Thus, following $f_x$ by $f_y$, does not make sense in the test case.

To solve this situation, the following verification has to be performed during the selection of the Read operations during the construction of the test case:

Let the test step $s_x = (f_x, (\text{'U'}, A_x))$ perform an Update operation on a data entity $e$, the function $f_x$ performs an U operation on entity $e$ and updates a set of attributes $A_x \subset A$. The test step $s_x$ has to be followed by a test step step $s_y = (f_y, (\text{'R'}, A_y))$, where $A_x \cap A_y$ is the largest possible. This applies for all Read operations following the Update operation by the coverage criteria defined in Section 3.2.2.

## 4. EVALUATION

To evaluate the efficiency of the proposed EDCyT, we compared it with common version of DCyT as defined in TMap Next [3] using the experiment described in this section.

For comparison of test cases produced by DCyT and EDCyT, we designed the experiment to answer the following principal questions. When test cases (consisting of test steps) are produced by DCyT and EDCyT for equivalent level of test coverage:

**Q1.** How many test steps in total are produced by DCyT and EDCyT?

**Q2.** How many inconsistent test steps are produced by DCyT and EDCyT?

**Q3.** How many data consistency defects remains undetected using the test cases produced by DCyT and EDCyT?

## 4.1 Experiment Method

Although we have collected feedback from actual applications of EDCyT that were conducted at several projects, EDCyT efficiency has been difficult to measure exactly due to the low quality of the test basis, continuous change requests and limited project schedules. Additionally, an exact measurement of the potential undetected defects of an actual project is difficult because these defects have been continuously fixed during the development and testing phase. For this reason, we have decided to simulate the test design process by DCyT and EDCyT on the model of an artificial application, which was semi-automatically generated for this purpose.

During the experiment, we created four instances of an artificial SUT model (we use the term instance of artificial SUT further on). The artificial SUT consists of functions, the states of the system, possible transitions between the states by the functions, the data entities handled by the system, the C, R, U, and D operations performed by the functions and other information, which are described in detail below. Also, a set of artificial defects were defined in particular artificial SUT instances.

Then, to verify the test cases, a group of selected test designers prepared test cases for a particular instance of the artificial SUT. First, the test cases were prepared by the DCyT technique, and then, they were prepared by several coverage variants of the proposed EDCyT technique. The test designers prepared the test cases with knowledge of the specifications of the artificial SUT. Information about the defined defects were not visible to the experimental group.

The test cases that were produced were then simulated in the model of artificial SUT to verify their consistency and the potential defects which remained undetected.

The experimental group was composed of 12 test designers, who varied by their previous praxis in software testing area from 1 year to 8 years. Majority of the experimental group members have background in finance, telecommunications and web enterprise software development. Approximately one third of the group members has systematically used DCyT previously. Approximately one half of the group has known the DCyT method. The DCyT method has been explained to the participants at the beginning of the experiment. Nobody from the group has been familiar with EDCyT before the experiment. At the beginning of the experiment, DCyT and EDCyT has been explained to participants.

### 4.1.1 Artificial SUT definition

Let consider an instance of an artificial SUT $A$ a tuple $(F, E, D, W, L, I)$; where,

$F$ is a set of SUT functions;

$E$ is a set of data entities used by the functions;

$S$ is a set of possible states of the SUT. The SUT changes its state when a function $f \in F$ is executed.

$D$ is a set of inserted data consistency defects. An inserted data consistency defect is a tuple $d = (e, f_c, o_c, F_d)$; where,

$e \in E$ is a data entity, which is in an inconsistent state that causes a defect;

$f_c \in F$ is the function that causes the data entity $e$ to be inconsistent;

$o_c \in \{C, U\}$ is the particular operation that causes the data entity $e$ to be inconsistent when accessed by the function $f_c$;

$F_d$ is a set of pairs ($f_d$, $o_d$); where,

$f_d \in F$ is a function that activates a defect in the SUT as a result of the inconsistency of the data entity $e$, which is caused by function $f_c$;

$o_d \in \{C, R, U, D\}$ is the particular operation that activates the defect in the SUT when it is accessed by function $f_d$.

$W$ is a set of workflows implemented in the SUT. The workflows describe the possible sequences of functions $f \in F$ in the SUT. The workflow is an oriented graph ($S_w$, $F_w$), whose nodes $S_w \in S$ are the states of the SUT, and its edges $F_w \in F$ are the functions of the SUT.

There can exist workflows $w_1$ and $w_2$, $w_1 \in W$, $w_2 \in W$, where $F_{w1} \cap F_{w2} = \varnothing$; where, $F_{w1}$ is a set of edges for workflow $w_1$, and $F_{w2}$ is a set of edges for workflow $w_2$.

$L$ is a set of data entity lifecycles in the SUT. For each data entity $e \in E$, the data lifecycle $l_e \in L$ is defined. The data entity lifecycle is an oriented graph ($S_l$, $F_l$), whose nodes $S_l \in S$ are states of the SUT, and whose edges $F_l \in F$ are functions of the SUT. $|L| = |E|$. For each of these edges, the C, R, U, D operations that are performed on an entity $e$ are defined.

$I$ is a set of inserted operations Influenced (I). The inserted operation Influenced is a triplet $i = (e_1, f_i, e_2)$, where function $f_i$ uses the data entity $e_2 \in E$ through one or more of the C, U, and D operations, and the particular data content of entity $e_1$ is changed by the function $f_i$, which primarily uses entity $e_2$.

To simulate the potential impact of the operation Influenced on the consistency of test case $c$, which was created for entity $e_1$, a probability is defined, for which the function $f_i$ causes the test case $c$ to be inconsistent.

As already defined in description of DCyT, test case $c$ for data entity $e$ is a sequence of steps $\{s_1, .., s_n\}$, where each of these steps is a pair ($f_s$, $o_s$), and $f_s \in F$, $o_s \in \{C, R, U, D\}$ is an operation that is performed on the data entity $e$ by the function $f_s$. Possible sequences of the SUT functions in the test cases are determined by $W$ and $L$.

Using the structure defined above, we are able to determine exactly which inserted defects can be detected by the prepared test cases and which defects would remain undetected. This analysis allows us to evaluate the simulation using some exact criteria that will be introduced further on in our discussion.

The criteria for an **inconsistent test step** $s_1$ are the following:

$s_2$ is the test step following $s_1$ in the test case sequence.

In the test steps $s_1$ and $s_2$, C, R, U, D or B operations are performed by functions $f_1 \in F$, $f_1 \in s_1$ and $f_2 \in F$, $f_2 \in s_2$.

Test step $s_1$ is inconsistent when two subsequent C, R, U, D, B operations, which are performed by the functions $f_1 \in F$ and $f_2 \in F$ in the test case, cannot be performed in the SUT; this occurs because we cannot reach a proper state in the SUT to execute the function $f_2$ from the SUT state reached by function $f_1$.

The criteria for **defect leakage** (defect which remains undetected by the test cases) are the following:

The data consistency defect $d \in D$ leaks when, for all $f_d \in F_d \in d$, the function $f_c \in d$ is either not present, or it is not followed by an $f_d$ in any of the test cases that were created for a particular instance of artificial SUT.

## 4.2 Experiment Results

For comparison of test cases for dynamic testing produced by DCyT and EDCyT, we prepared four instances of an artificial SUT that differed by the $F$, $E$, $D$, $W$, $L$, and $I$. These instances were randomly distributed to the test designers, simulating the test basis (the design documentation, from which the test cases were created). Presence of the artificial defects has been concealed to the experiment participants. In the test design process, the test designers created CRUD matrix from the given test basis.

Then, every participant created his or her own test cases for assigned artificial SUT instances by the following methods:

The test designers first used the TMap definition of DCyT without any other enhancements of EDCyT as proposed in this paper.

Because the coverage criteria are not explicitly defined in TMap, the test designers were instructed to use the following description, which was included in the methodology: "After every action (C, U or D), an R is carried once. With respect to the relevant entity, all occurrences of the actions C, R, U, and D in all the functions should be covered" (further DCyT 1R), and "More thorough coverage can be achieved by requiring that the combinations of actions are also fully covered. For example, this coverage can be achieved by requiring that after each U all the functions with R should be carried out" (further DCyT NR).

The test designers prepared the test cases using the DCyT method with the coverage criteria DCyT 1R and DCyT NR for each of the artificial SUT instances. Then, the produced test cases were taken from the designers. Then, the designers were introduced to our proposal of the EDCyT. After clarification of some initial questions, the test designers prepared test cases using EDCyT by the coverage criteria (sequentially) 1R, NR, 1RI and NRI for each of the artificial SUT instances. In this phase, each test designer was provided with a different artificial SUT instance than in the previous DCyT phase.

Then, the produced test cases were taken away from the test designers. Finally, the collected test cases were simulated in an artificial SUT to assess inconsistent test case steps and ratio of artificial defects, which would remain undetected by the test cases (defect leakage).

In the experiment, we compared coverage DCyT 1R with EDCyT 1R and 1RI, and we also compared DCyT NR with EDCyT NR and NRI.

**Table 1. Averaged results for all artificial SUT instances**

| Method and coverage criteria | Averaged inconsistent test step ratio | Averaged defect leakage ratio | *Test steps increase* |
|---|---|---|---|
| DCyT 1R | 13.0% | 22.0% | - |
| EDCyT 1R | 9.5% | 16.9% | 6.4% |

| EDCyT 1RI | 5.2% | 12.0% | 19.0% |
| DCyT NR | 11.8% | 11.5% | - |
| EDCyT NR | 5.3% | 7.9% | 12.7% |
| EDCyT NRI | 3.5% | 6.4% | 17.5% |

The overall results are provided in Table 1. We averaged the ratio of inconsistent test case steps and the defect leakage ratio for all instances of the artificial SUT. Additionally, we provide the averaged increase in the number of test steps that was produced by the proposed extension.

In Table 1, *Test steps increase* stands for Averaged increase in number of EDCyT test steps compared to DCyT.

We compared test cases produced by DCyT and EDCyT for equivalent coverage criteria. The results for EDCyT show that, in synergy, (1) the proposed test coverage criteria including an exact method for test case creation, (2) the operation Influenced, and (3) the proposed approach for verification of the consistency of the test cases lead to higher number of test steps in produced test cases (Q1), but lower numbers of inconsistent test steps, which would have been re-evaluated during an actual testing process and this would cause an additional overhead to the project (Q2), and less data consistency defects which would remain undetected using the produced test cases (Q3).

## 5. RELATED WORK

Common definition of the The Data Cycle Test (DCyT) [3,5-7] usually consists of (1) a guideline of how to create a CRUD matrix, (2) several high-level rules of possible static testing using a CRUD matrix, and (3) a guideline how to create the dynamic test cases as a sequences of Create, Update, Read and Delete operations affecting a data entity. In TMap Next [3], master and derived data, which influence the test design process, are discussed briefly. Nevertheless, this concept does not cover all possible situations of data dependencies and data influencing, which can be implemented in SUT.

From the related areas, the data-flow testing is conceptually the closest to the subject of this paper. The goal of the proposed EDCyT has not been directly discussed in the related literature, however, related previous results from the data-flow testing area shall be analyzed. Let us start this overview by a static testing area, then we will comment on previous extensions of the CRUD Matrix used for this purpose and, finally, we summarize the dynamic testing area.

The data flow analysis works focus on detection of design errors in workflow design [8-10]; also detection and automated correction measures are proposed [11]. These works principally relate to static testing using different notations than the CRUD Matrix. UML activity diagrams [9], UML state-chart diagrams [10], Petri's nets [8,11] or IFML diagrams [12] are used as an SUT model for instance. Alternatives to established formal data flow analysis can also be found, attempts have been made to combine model-based testing in this field with exploratory testing [13,14], to construct a model of the system under test dynamically during the testing process.

Extensions of the CRUD Matrix by attributes has been already proposed by Jukic et al. [15,16]. However, in this proposal, attributes are added as an extra column to the CRUD Matrix [15], which is different to our approach. For static testing of database design, an extension of the CRUD matrix by individual operations on the physical design level has been also proposed [17].

Regarding the dynamic testing area, the Data Flow Analysis techniques [18-20] seems similar to the DCyT and EDCyT. However, there are principal differences.

The Data Flow Analysis is white-box testing technique, which is using a program source code as the test basis. As the data objects, particular values of the program variables are analyzed. To design the test data flow cases, the set-use pairs are defined for three principal actions: (1) definition of the variable (initialization and setting of an initial value of a variable), (2) use of the variable (reading the actual value of the variable) and (3) destruction of the variable. Then, in the data flow technique, we analyze, which combinations of the pairs indicate possible defects in the tested source code.

Principally, the set-use information is not captured by the CRUD Matrix, which serves as a model of the SUT in the DCyT and EDCyT. These techniques operate on a higher level of abstraction – a conceptual design level of the SUT.

The DCyT and EDCyT use general conceptual data entities whose data a modified by the SUT functions. Usually, the DCyT test cases are used for the black-box functional testing, based on testers' interaction with the user interface of the SUT.

Also, alternative modeling notations to the CRUD Matrix can be found. The Data-Flow Matrices [21] contain Read and Write operations performed on data objects by SUT functions. Differently, to the CRUD Matrix, this proposal defines Read and Write operations only. In this approach, data flow information is combined with a workflow model to detect possible missing, redundant or conflicting data. In contrast, EDCyT proposes an extended CRUD Matrix, which is used to define dynamic test cases. These test cases focus on testing of consistency of data entities used by SUT processes.

Dependencies between SUT data entities, which are the subject of our research in EDCyT, have been previously addressed by [22] for example. The authors question the coverage criteria that were used for previous definitions of the data flow test techniques, and they propose an extension to this coverage that is based on introducing data dependencies. Nevertheless, this approach is based on Linear Temporal Logic and flow graphs, which are not directly applicable to the scope of our investigation because we assume the availability of the CRUD matrix only. The same situation applies to approaches that are based on state machines; for example presented by [23], or it may be based on a combination of various types of UML diagrams that were gathered from the test basis such as statechart and sequence diagrams as described by [24].

Apart from the field of dynamic testing, SUT model based on CRUD matrices has also been employed for static testing technique [25]; however, the principle of these tests differ from the approach presented in this study.

## 6. CONCLUSION

In this paper, we presented an extension of the Data Cycle Test (DCyT) [3] technique, the EDCyT. The main use of EDCyT is intense testing of priority functions or modules of the SUT, where data consistency is an important factor for correct system behavior. In EDCyT, we proposed extensions with the goal of increasing the consistency of the resulting test cases; we also defined test coverage criteria for the resulting test cases, and we provided the test designer with a more exact method of how to chain C, R, U, D operations to a test case. For this reason, we proposed two more operations of the CRUD matrix: I (Influenced) and B (Best read).

The experimental results described here show that, with an EDCyT extension, the ratio of potentially inconsistent test case steps to the potential number of undetected data consistency defects is lower. Using the EDCyT, the number of test case steps is higher, which is the cost of better test coverage. During the experiments, more than 80% of the test designers involved in the experiment rated the EDCyT highly for (1) providing test coverage that was more controlled, (2) providing more convenience regarding the exact rules used to define test cases, (3) providing test cases that were more consistent. During the experiments, we incorporated several practical comments into the proposed method.

To precisely measure the efficiency of the produced test cases and to keep the state of the SUT fixed, we verified EDCyT using a test design experiment with the help of a group of test designers who used several instances of the model of the artificial SUT. In an actual project, the result could differ with the internal structure of the SUT, the number of influencing data entities and the number of data consistency defects in the SUT. Nevertheless, the results from experiments using artificial SUT are indicative of good performance of the technique and motivate us to evolve the proposed EDCyT technique further.

## 7. ACKNOWLEDGMENTS

This study is conducted as a part of the project TACR TH02010296 „Quality Assurance for Internet of Things Technology". This work has been supported by the OP VVV funded project CZ.02.1.01/0.0./0.0./16_019/0000765 „Research Center for Informatics".